\documentclass[article]{elsarticle}

\usepackage{graphicx}
\usepackage{dcolumn}
\usepackage{bm}
\usepackage{subfig}
\usepackage{tikz}
\usepackage{color}
\usepackage{float}
\usepackage{amsfonts}
\usepackage{amsmath}

\newtheorem{definition}{Definition}
\usepackage{lineno,hyperref}

\journal{IOP: FDR}
\begin{document}

\begin{frontmatter}

\title{Lagrangian angular momentum vortex definition and tracking}

\author{Anass El Aouni$^{1}$ and Arthur Vidard }
\address{Univ. Grenoble Alpes, Inria, CNRS, Grenoble INP, LJK, 38000 Grenoble, France}
\fntext[myfootnote]{anas.elaouni@gmail.com}

\begin{abstract}
In the present paper, we study transport properties of coherent vortices. These structures are formed by tubes of fluid parcels that complete similar material rotation.
Here, we demonstrate that time $t_0$ positions of such physical structures can be identified from the angular momentum along their trajectories.
We identify and extract these Lagrangian vortices as material domain in which fluid elements complete similar Lagrangian angular momentum around the same axis and over the same finite time duration.  The present method shows high vortices' monitoring capacity by identifying automatically all coherent clockwise and counterclockwise vortices.  The performance of the proposed method is illustrated over different two- and three-dimensional flows.
\end{abstract}


\end{frontmatter}


\section{Introduction}

Vortices are omnipresent in our world, they can be observed with different scales both in time and size. As for now, a universal definition of coherent vortices  is deceptively complicated and remains an open issue in fluid mechanics, which is likely one of the main complication behind substantial confusion in turbulence research. However, a main feature of a possible definition is broadly agreed;  vortices are intuitively recognized as evolving domains with rotational motion of the fluids.

In classical vortex dynamics, vortices are usually associated with vorticity. Study in \cite{lamb1924hydrodynamics} defines vortices from vorticity tubes. Authors in \cite{wu2007vorticity} defines vortices as regions of high concentration of vorticity compared to its surrounding. Authors in \cite{elsas2017vortex} propose a vorticity curvature criterion method to identify coherent vortices.
Such vorticity-based methods are simple, however, they are not always satisfying. As an example of contradiction, the vorticity cannot distinguish a vortical region from a shear layer region, therefore, regions with high vorticity concentration but no rotation motion are considered as vortices.

Along with the vorticity-based methods, several other definitions have been proposed in the literature along with their numerical methods, both in the Eulerian and Lagrangian perspectives. An $\omega_R$ criterion to identify vortices has been developed in \cite{dong2019new,liu2016new}. These work by computing the strength of the relative rotation on the plane perpendicular to the local rotation axis. Authors in \cite{liu2018rortex} consider a vortex as an homogeneous region characterized by a magnitude of vortex vector larger than zero.
The work in \cite{perry1987annu} defines vortices as regions where eigenvalues of velocity gradient tensor are complex.
Authors in \cite{zhou1999mechanisms} proposed using iso-surfaces of the eigenvalue's imaginary part to identify vortices.
Another well-known method was developed in \cite{jeong1995identification}, the authors used the second eigenvalue of the symmetric component of the gradient tensor to capture vortices.
Based on the spinning movement of parcels withing vortices, authors in \cite{el2019fourier} proposed an approach which defines vortices from the frequency-domain representation of their observed trajectories, it defines them as closed material surfaces along which fluid elements share similar frequency components.
Authors in \cite{el2020defining} defines vortices from their observed Lagrangian trajectories. This has been done by decomposing particles trajectories into two parts:  a part which describes the mean displacement and closed curves which give information about uniformly rotating flow.
Other studies used probabilistic approaches which use the evolution of probability densities and almost-invariant sets \cite{froyland2013analytic,froyland2012three}.

The method proposed in the present paper is based on the analysis of the Lagrangian velocity vectors along particles' trajectory. Angular momentum of a particle exhibiting rotation oscillates around a centerline with a given frequency which presents the vortex-turn-over time. This oscillation between negative and positive velocity vector with regards to its centerline yields a consistent measure of material rotation and direction.
We define Lagrangian coherent vortex as closed material surfaces in which fluid parcels complete similar rotation. This turns out to be filled with tubular level-sets of the Lagrangian angular momentum along particles' trajectory.
The rest of this paper is organized as follows: Section II describes the setup and outlines the main computational tool.
Section III discusses and illustrates particles' velocities within vortex.
Section IV presents and details our new approach. Section V illustrate our method via different fluid simulations. Conclusion is drawn in the last Section.

\section{ Set-up}

We consider a time-dependent smooth vector field:
\begin{equation}
  \mathbf{v}(\mathbf{x},t), \quad \mathbf{x} \in \mathbb{R}^3, \quad t \in [\alpha,\beta]
\end{equation}

and its associated ordinary differential equation:

\begin{equation}\label{eq:ode1}
  \dot{\mathbf{x}}=\mathbf{v}(\mathbf{x},t), \quad \mathbf{x} \in \mathbb{R}^3, \quad t \in [\alpha,\beta]
\end{equation}

where $\mathbf{v}$ a smooth velocity field defined on a domain:

\begin{equation}
\mathbf{U}(t)  \subset \mathbb{R}^3, \quad \mathbf{U}= \bigcup_{ t \in [\alpha,\beta]}  \mathbf{U}(t) \subset \mathbb{R}^3 \times [\alpha, \beta]
\end{equation}

The flow map is defined as the map that takes a particle from its initial location $\mathbf{x}_0$ at time $t_0$ to its  location $\mathbf{x}_t$ at time $t$:

\begin{equation}\label{eq:2}
  \mathbf{F}^{t}_{t_0}(\mathbf{x}_0):=\mathbf{x}(t,t_0,\mathbf{x}_0), \quad \alpha \leq t_0 \leq t \leq \beta,
\end{equation}

$\mathbf{x}(t,t_0,\mathbf{x}_0)$ denoting the trajectory of Eq.\ref{eq:ode1} passing through a point $\mathbf{x}_0$ at time $t_0$.

Consider a  material domain (defined by a set of fluid particles) $\mathcal{M}(t_0)$ advected by the flow. Its image at time $t$ can be expressed in term of the flow map as $\mathcal{M}(t)=\mathbf{F}^{t}_{t_0}(\mathcal{M}(t_0))$.

\section{Trajectories and velocity vectors of fluid parcels withing vortex}

\begin{figure}[h]
  \centering

\includegraphics[width=0.5\textwidth]{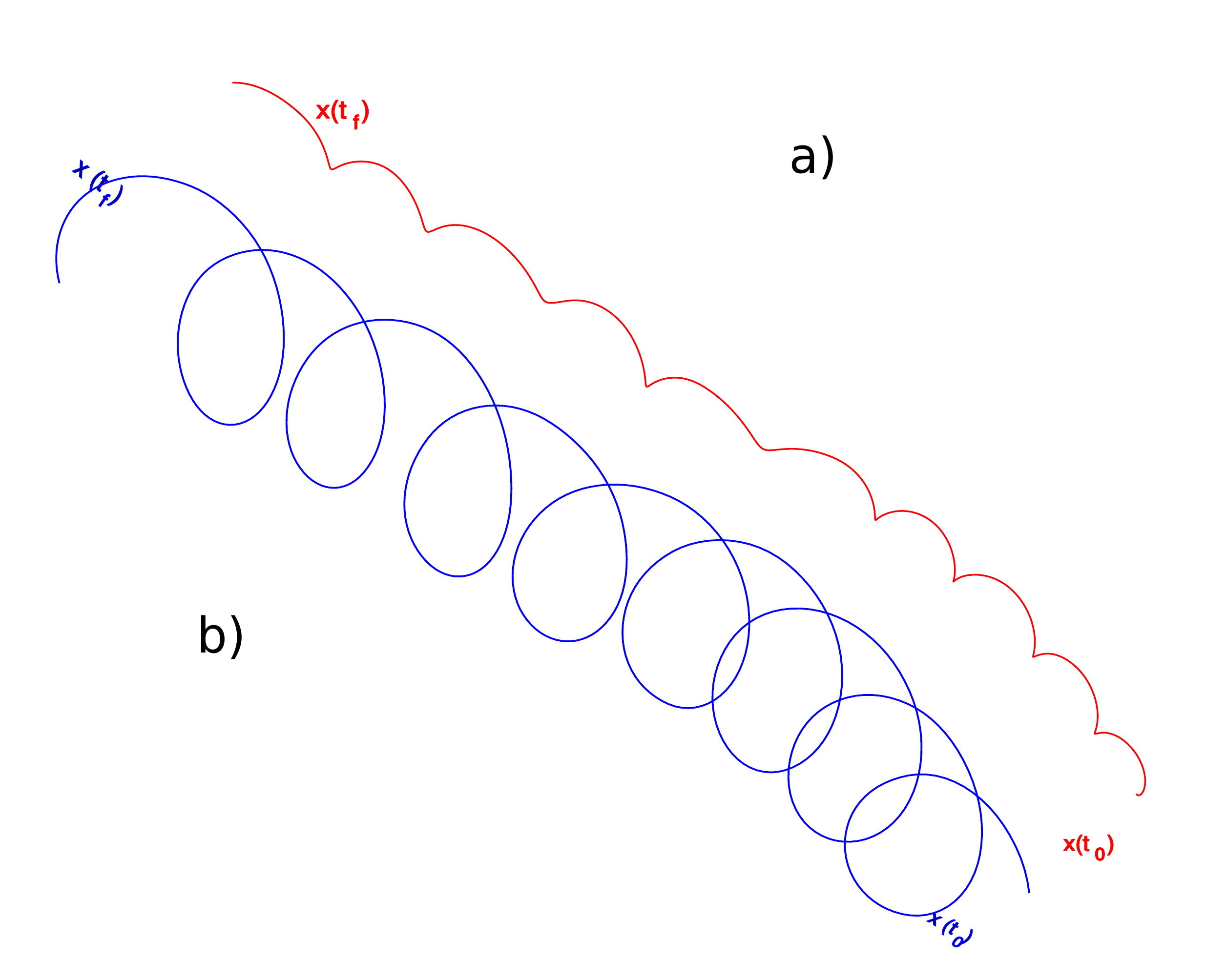}

\caption{ Two particles trajectories initialized within different vortices characterized by difference angular velocities; the blue one crosses itself and create closed curves, while the red one does not.\label{fig:1}}
\end{figure}

Trajectory of particle withing a vortex results  into a loopy curve which intersect itself and creates closed curves as shown in Fig.\ref{fig:1}(b). Based on this conceptual view, authors in \cite{dong2011scheme} proposed a scheme to identify loops from trajectories of oceanic surface drifters. Based on the same concept, authors in \cite{el2020defining} proposed the Averaged Closed Curve metric to identify Lagrangian vortices. Such methods provide satisfying results when particles trajectories  intersect themselves.
However, a particle within a vortex might not necessary intersect its trajectory as we show in Fig.\ref{fig:1}(a); this figure  shows a trajectory initialized within a vortex characterized by different angular momentum where no closed-curve is recorded. Another example is a vortex with zero-translating speed and radial flow where the particles' trajectories converge according to inflow; a trajectory of a particle within such vortex will not intersect itself. 

\begin{figure}[h]
  \centering

\includegraphics[width=0.5\textwidth]{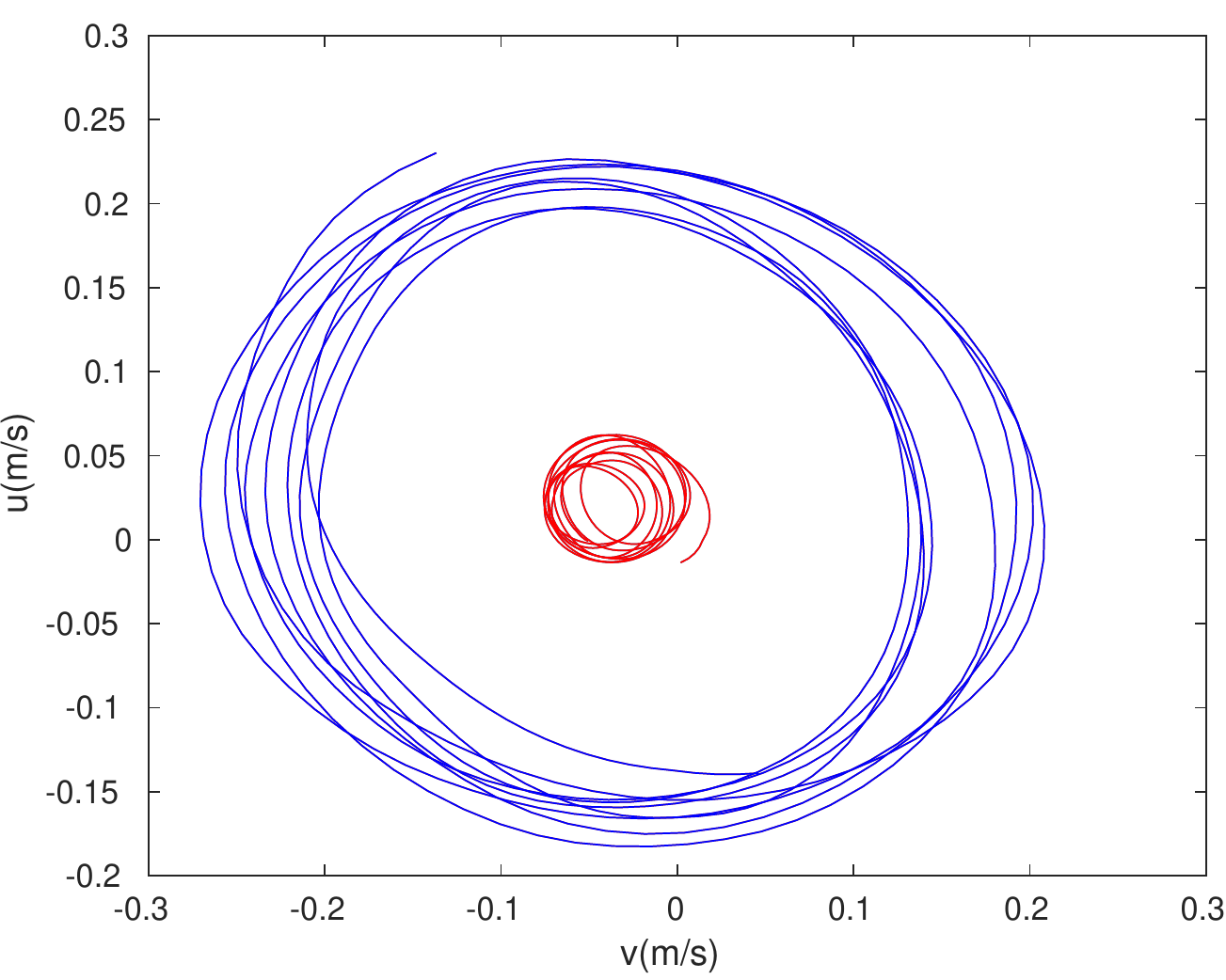}

\caption{ Lagrangian angular momentum along the trajectories in Fig.\ref{fig:1}. (See the supplemental movie M0 for the time evolution of their vectors.)\label{fig:2}}
\end{figure}

We have shown that focusing on the analysis of particle's trajectory leaves several unaddressed problems.
On the other side, the Lagrangian angular momentum along these trajectories gives an insight view of their dynamics. We show in Fig.\ref{fig:2} the Lagrangian angular momentum along the trajectories in Fig.\ref{fig:1}. We see that while the trajectories in Fig.\ref{fig:1} show different patterns, their Lagrangian angular momentum show similar one. It shows that both particles undergo rotational movement; each of these Lagrangian angular momentum oscillates between negative and positive phases with regards to its centerline. We show in the supplemental movie M0 the time evolution of the Lagrangian angular momentum along the trajectories in Fig.\ref{fig:1}.

\section{Defining coherent vortex from velocity vectors along its Lagrangian trajectory}

Here, we aim to define Lagrangian coherent vortex from the angular momentum along its trajectory. We seek to identify this structures as closed material surface along which fluid parcels exhibit similar rotations around the same axis, and over the same finite time interval.
We have shown in the previous section that particles within vortices might not necessarily have similar trajectories' patterns but rather similar Lagrangian angular momentum patterns. This latter oscillate between positive and negative phases with regard to a given centerline, and with a given frequency representing the vortex-turn-over time.
Based on this, we define the Lagrangian Angular Momentum $(\mathcal{LAM})$ along particles' trajectory as: 

\begin{equation}
 \mathcal{LAM}(\mathbf{x}(t_f,t_0,\mathbf{x}_0))=\int_{t_0}^{t_f} \omega(\mathbf{x}_0,t) dt
\end{equation}

Where $\mathbf{\omega}(\mathbf{x}_0,t)$ presents the Lagrangian angular velocity of the particle $\mathbf{x_0}$. This can be expressed as:

\begin{equation}
 \mathcal{LAM}(\mathbf{x}(t_f,t_0,\mathbf{x}_0))=\sum_{i=1}^{n-1} \cos^{-1} \frac{ \vec{{\mathbf{v}}}(\mathbf{x}_0,{t_i}) \cdot \vec{{\mathbf{v}}}(\mathbf{x}_0,{t_{i+1}})}{\left | \vec{{\mathbf{v}}}(\mathbf{x}_0,{t_i}) \right |  \left | \vec{{\mathbf{v}}}(\mathbf{x}_0,{t_{i+1})} \right |}
\end{equation}

\begin{figure}[h]
  \centering

\includegraphics[width=0.5\textwidth]{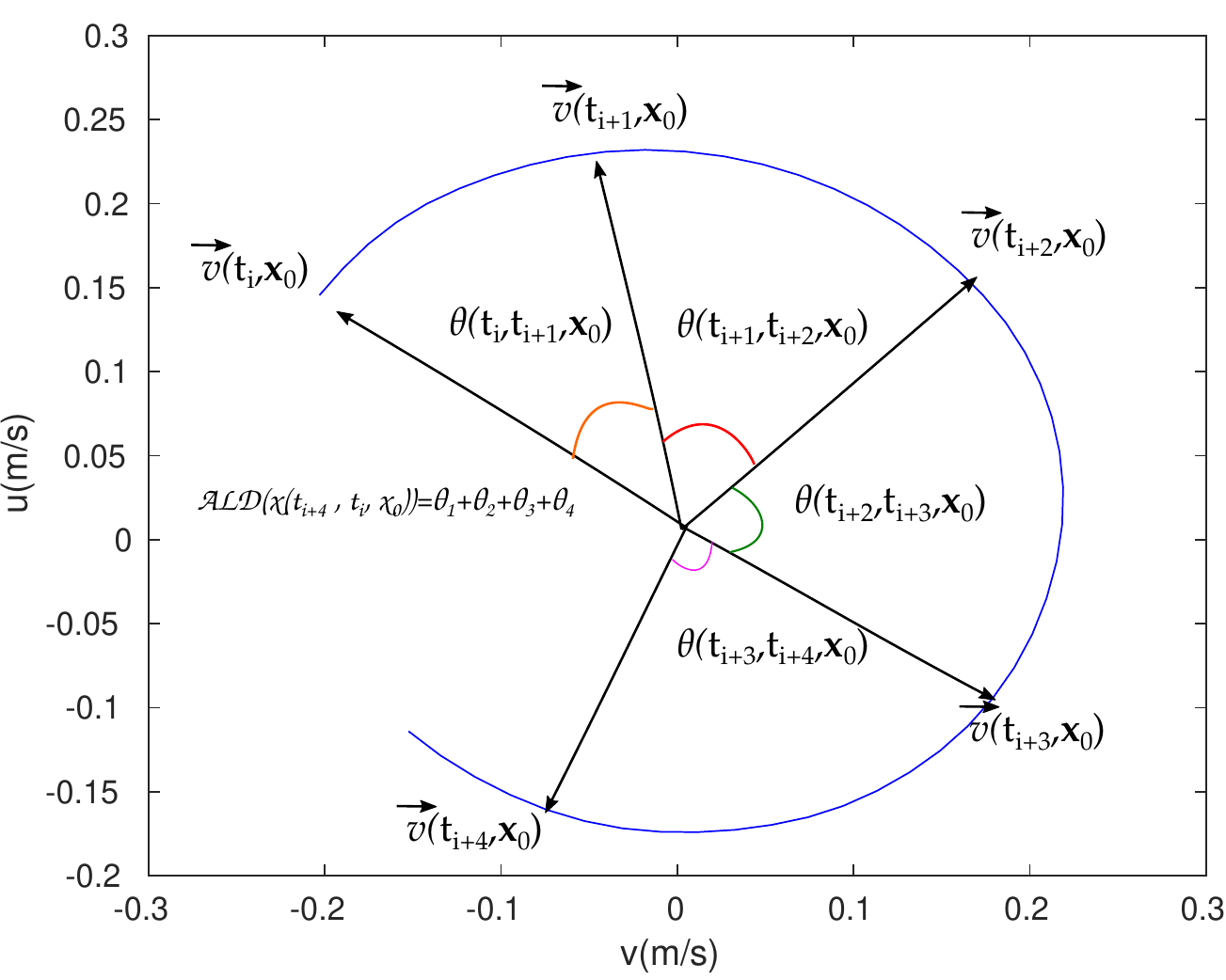}

\caption{ Geometrical view of $\mathcal{LAM}$ calculation over a segment of the Lagrangian velocity vector along the trajectory in Fig.\ref{fig:1}a.\label{fig:3}}
\end{figure}

Where $\frac{\cos^{-1} \vec{{\mathbf{v}}}(\mathbf{x}_0,{t_i}) \cdot \vec{{\mathbf{v}}}(\mathbf{x}_0,{t_{i+1}})}{\left | \vec{{\mathbf{v}}}(\mathbf{x}_0,{t_i}) \right |  \left | \vec{{\mathbf{v}}}(\mathbf{x}_0,{t_{i+1})} \right |}$  represents the angle $\mathbf{\theta}(t_{i},t_{i+1},\mathbf{x}_0)$ between two successive velocities' vectors $\left \{ \vec{\mathbf{v}}(t_{i},\mathbf{x}_0),\vec{\mathbf{v}}(t_{i+1},\mathbf{x}_0) \right \}$ of the Lagrangian velocity $\vec{\mathbf{v}}(t_f,t_0,\mathbf{x}_0)$.
This represents the Lagrangian total angle accomplished by the velocity vectors along a given trajectory, which gives a consistent measure of particle rotation. Along with the total angle, the $\mathcal{LAM}_{t_f}^{t_0}$ distinguishes between clockwise and counterclockwise vortices.
We show in Fig.\ref{fig:3} the geometrical view of the $\mathcal{LAM}$ method applied on a segment of the Lagrangian velocity vector of particle within a vortex.

We now use the $\mathcal{LAM}$ to identify closed material surface along which fluid parcels experience similar rotations over the same time interval $[t_0 ~ t_f]$.
Time $t_0$ positions of such material tubes are tubular level ( a contour in two dimensions, and a cylindrical in three dimensions) surfaces of the scalar  $\mathcal{LAM}_{t_f}^{t_0}$. 
By definition, the $\mathcal{LAM}_{t_f}^{t_0}$ map allows the identification of vortex' center as inner most member of $\mathcal{LAM}_{t_f}^{t_0}$ within the vortex.

\begin{figure}[h]
  \centering

\includegraphics[width=0.5\textwidth]{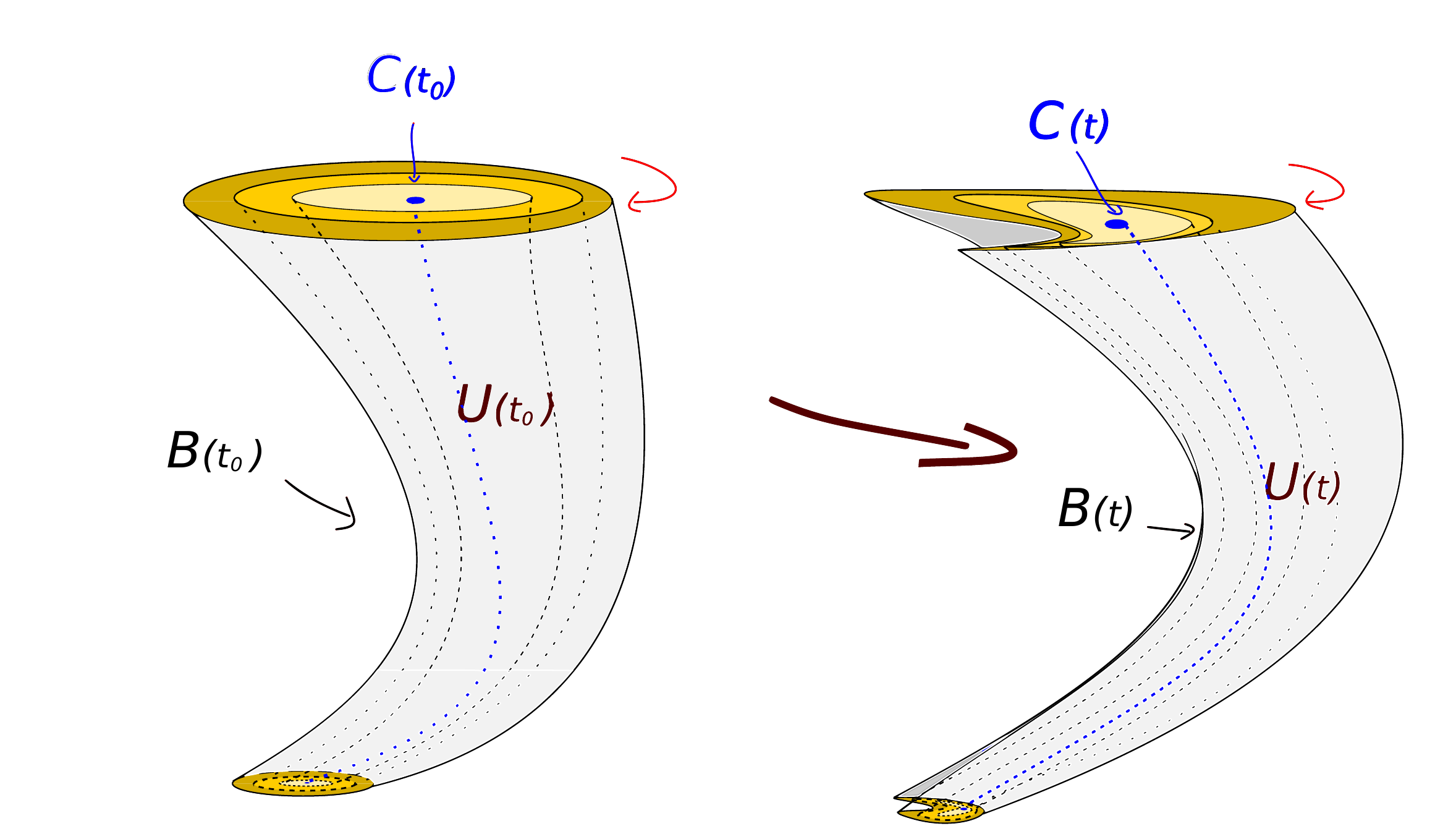}

\caption{ Time $t_0$ and $t$ positions of a coherent Lagrangian vortex $\mathbf{U}(t)$, its boundary $\mathbf{B}(t)$ and center $\mathbf{C}(t)$.\label{fig:bis_draw}}
\end{figure}

We summarize the $\mathcal{LAM}$-based vortex identification in the following definition, with its geometry illustrated in Fig.\ref{fig:bis_draw}.

\begin{definition}For a given time interval $[ t_0, t_f ]$:
\begin{itemize}
  \item 1)  Lagrangian coherent vortex is an evolving material domain $\mathbf{U}= \bigcup_{ t \in [t_0,t_f]}  \mathbf{U}(t) \subset \mathbb{R}^3 \times [t_0, t_f]$ such that $\mathbf{U}(t_0) \subset \mathbb{R}^3$ is filled with a nested family of tubular level surfaces of $\mathcal{LAM}_{t_f}^{t_0}$.
  \item 2) The boundary $\mathbf{B}= \bigcup_{ t \in [t_0,t_f]}  \mathbf{B}(t) \subset \mathbb{R}^3 \times [t_0, t_f]$  of $\mathbf{U}$ is the outermost closed material surface of $\mathcal{LAM}_{t_f}^{t_0}$ in $\mathbf{U}(t_0)$. 
 \item 3) The center $\mathbf{C}= \bigcup_{ t \in [t_0,t_f]}  \mathbf{C}(t) \subset \mathbb{R}^3 \times [t_0, t_f]$ of $\mathbf{U}$ is defined as a material set $\mathbf{C}$ such that $\mathbf{C}(t_0)$ is  the innermost (minimum) member of the magnitude   $\mathcal{LAM}_{t_f}^{t_0}$.

 \item 4) The orientation of the vortex $\mathbf{U}(t_0) \subset \mathbb{R}^3$  is defined by the sign $sgn(\mathcal{LAM}_{t_f}^{t_0})$ in  $\mathbf{U}(t_0)$.
\end{itemize}
\end{definition}

Lagrangian vortex, its boundary and center are materials objects. Thus, their time $t$ positions are only determined by Lagrangian advection:

\begin{equation}\label{eq:8}
  \mathbf{U}(t)=\mathbf{F}^{t}_{t_0}(\mathbf{U}(t_0)), \quad \mathbf{B}(t)=\mathbf{F}^{t}_{t_0}(\mathbf{B}(t_0)), \quad \mathbf{C}(t)=\mathbf{F}^{t}_{t_0}(\mathbf{C}(t_0)), \\ \quad t \in [t_0, t_f]   \\
\end{equation}

The $\mathcal{LAM}$-vortex approach differs from the previous definitions, it does define vortex based on the total angle accomplished by the Lagrangian velocity vectors along their trajectories; this insure that all the detected particles exhibit rotational movements. $\mathcal{LAM}$-vortex might shows material filament, but by definition, they all rotate together with the vortex without breaking away.

\section{Experiments}

In this section, we apply our method to different fluid simulations to confirm our theoretical part.

\subsection{Two-dimensional examples}

\subsubsection{Direct numerical simulation of two-dimensional turbulence}

We start by solving numerically the Navier-Stokes PDE equation for a two-dimensional velocity, $u: \mathcal{D} \to \mathbb{R}^2 $ of an incompressible fluid on periodic domain, $\mathcal{D} = [0, 1] \times[0, 1]$, this is expressed  as:

\begin{align} 
\partial u_t+u \cdot \nabla u &=-\nabla p+\frac{1}{Re} \Delta u+f,&  (x,t) \in \mathcal{D} \times [a,b],\\
\nabla \cdot  u &= 0,&  (x,t) \in \mathcal{D} \times [a,b],\\
\int u^{j}dx &= 0,&  (x,t) \in \mathcal{D} \times [a,b], j=1,2,\\
u &= u^{*},&  (x,t) \in \mathcal{D} \times  \{0\} \label{eq:nv1}
\end{align} 

Here $u \cdot \nabla u$ presents the inertial term which is responsible for the transfer of kinetic energy in the turbulent cascade. $\frac{1}{Re} \Delta u$ represents the dissipative viscous term, and $\nabla p$ is the pressure gradients which insures the incompressibility of the flow.
We further assume periodic boundary conditions and  use a standard pseudo-spectral method with 128 modes in each direction and $2/3$ dealiasing to solve the above Navier-Stokes model with Reynolds number $Re = 10^{4}$ on the time interval $t \in [0, 600]$.  The model is parametrized by the pressure function $p : D\times[a, b] \rightarrow \mathbb{R}$, with no external forcing ($f=0$). 
 We initialize the system with the vorticity of two adjacent vortices:
\begin{equation}
\overline{\omega}|_{t_0}=exp(\frac{(x+0.7)^2+(y-0.6)^2}{0.2}) - exp(\frac{(x+0.7)^2+(y-0.4)^2}{0.2})
\end{equation}

\begin{figure}[h]
  \centering
\includegraphics[width=.55\textwidth]{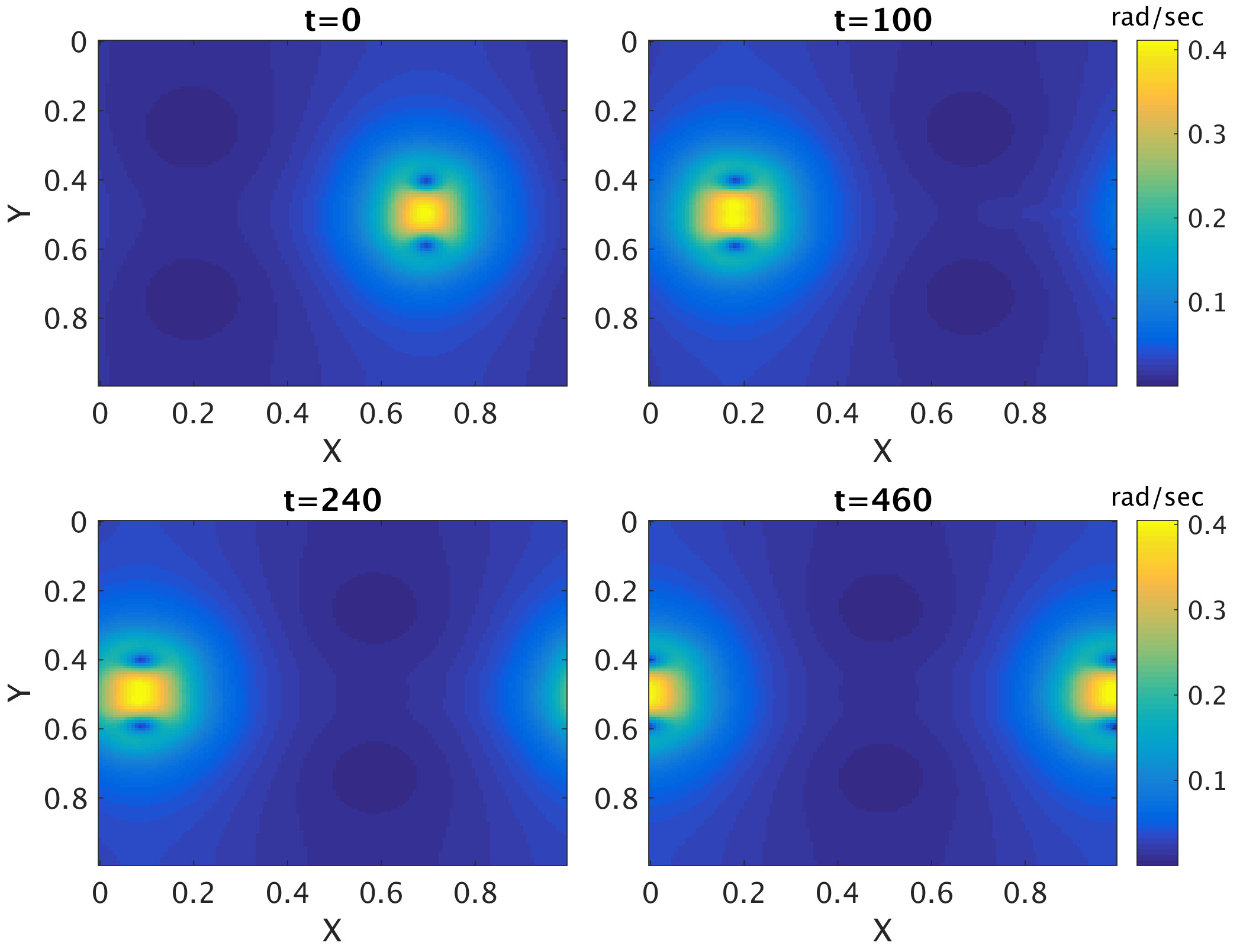}
\caption{Time evolution of the magnitude of the velocity governed by the Navier-Stokes model \ref{eq:nv1} evaluated at times t = (0, 100, 240, 460). \label{fig:bis4}}
\end{figure}

The vorticity stream formulation \cite{spotz1995high} is used to extract the velocity and pressure from the stream function.
We then carry out the flow integration over the interval $t \in [0, 600]$ by a fourth-order Runge-Kutta method with variable step-size.
Fig.\ref{fig:bis4}  displays the simulated two-dimensional velocity field of the Navier-Stokes model \ref{eq:nv1} captured at 4 different time points.

We construct the  $\mathcal{LAM}^{t_f}_{t_0}$  map by using an initial grid of $128\times128$ and integrate the two-dimensional simulated velocity field over the time interval $[t_0=0, t_f=600]$ using the eq.\ref{eq:ode1}.

\begin{figure}[]
  \centering
  \subfloat[]{\includegraphics[width=.7\textwidth]{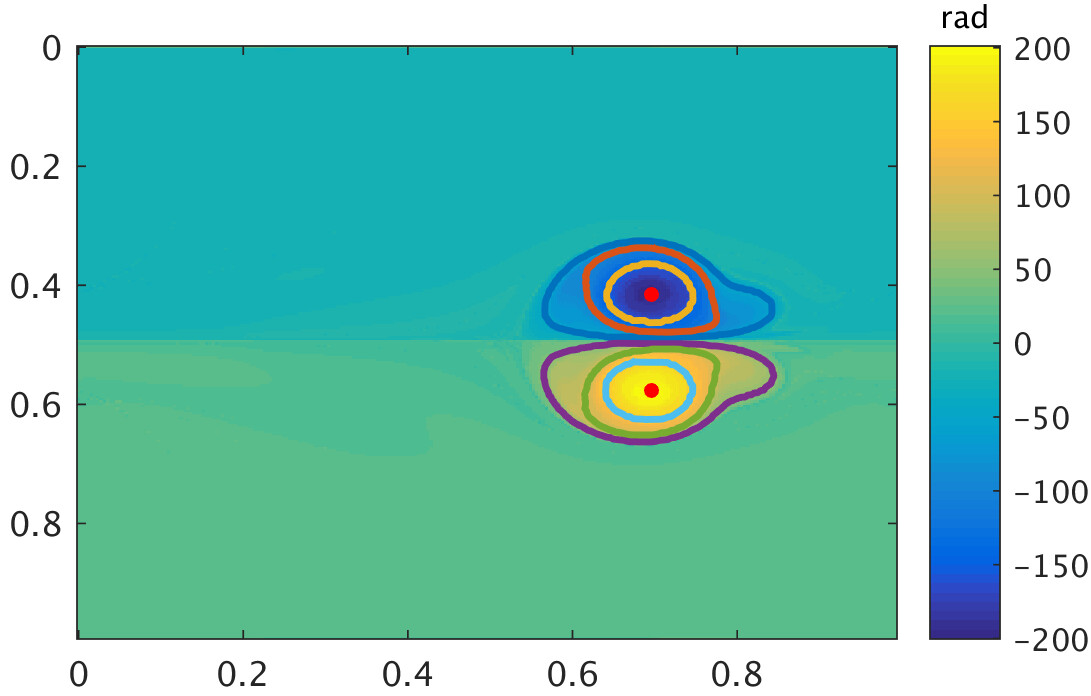}}
\subfloat[]{\includegraphics[width=0.4\textwidth]{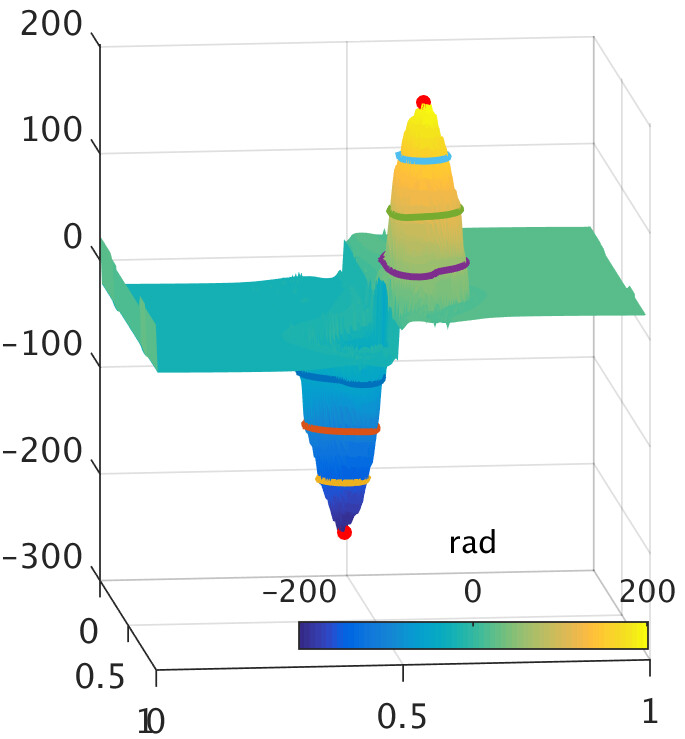}}

\subfloat[]{\includegraphics[width=0.8\textwidth]{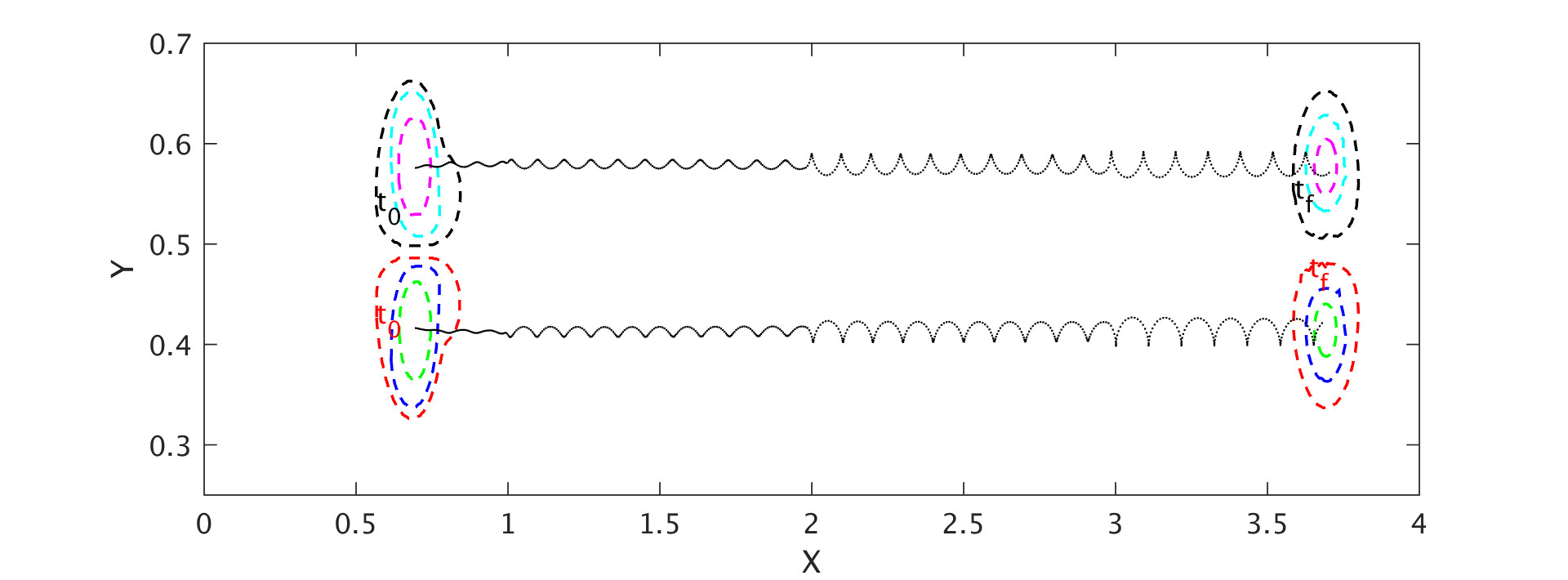}}

\caption{a) Coherent vortices at time $t_0$ extracted from the velocity field generated by Navier-Stokes model \ref{eq:nv1}  using definition.1 with the $\mathcal{LAM}_{t_f}^{t_0}$ map shown in background. b) three-dimentional contour plot of  $\mathcal{LAM}_{t_f}^{t_0}$, three vortex cores are extracted from each clockwise and counterclockwise vortex along with their centers. c) Their initial and final positions under Lagrangian advection as well as trajectories of their centers. (The supplemental movie M1  shows the complete advection sequence of these vortices. It also shows the oscillation of velocity vectors of different particles withing these two vortices.)\label{fig:bis5}}
\end{figure}

Fig.\ref{fig:bis5}(a) shows the boundaries of two coherent vortices and their centers extracted from the $\mathcal{LAM}^{t_f}_{t_0}$ map of the simulated velocity field. Within each vortex we extract different cores characterized by different total angle values. Along with the extracted vortices, the $\mathcal{LAM}_{t_f}^{t_0}$ map gives information about the orientation of the vortices; negative values refers to clockwise vortex while positive values refers to counterclockwise one.  
Fig.\ref{fig:bis5}(b) shows a three-dimentional contour plot of  $\mathcal{LAM}_{t_f}^{t_0}$, highlighting positions of different cores extracted from these clockwise and counterclockwise vortices.
Fig. \ref{fig:bis5}(c) shows their final position under Lagrangian advection and trajectories of their centers. These Lagrangian vortices  maintain their coherency, they do not stretch or fold. The time evolution of these vortices over the interval of advection $[t_0=0, t_f=600]$ is illustrated in the movie M1.

\subsection{Two-dimensional eddies in satellite geostrophic velocity}

In this part, we use sea surface current data to illustrate the automatic identification and extraction of coherent vortices. These data is derived from satellite altimetry under the geostrophic approximation where sea-surface height $\eta (\varphi,\theta, t)$ serves as a non-canonical Hamiltonian for surface velocities in the $ (\varphi,\theta)$ longitude-latitude coordinate system. The evolution of fluid particles satisfies:
\begin{align}
 \dot{\varphi}(\varphi,\theta,t)&=-\frac{g}{R^2f(\theta)\cos\theta}\partial_{\theta}\eta(\varphi,\theta,t)\\
 \dot{\theta}(\varphi,\theta,t)&=\frac{g}{R^2f(\theta)\cos\theta}\partial_{\varphi}\eta(\varphi,\theta,t) \label{eq:ssh}
\end{align}

 Here, $g$ presents the constant of gravity, $R$ the mean radius of the Earth and $f(\theta) = 2\Omega\sin\theta$  the Coriolis effect, with $\Omega$ denoting the Earth's mean angular velocity.
The data used here is produced by Ssalto/duacs multi-mission sea level products provided by AVISO \footnotemark[2] with a spatial resolution of $1/4^{\circ}$ and temporal resolution of 7 days.

\begin{figure}[ht]
  \centering
  \subfloat[]{ \includegraphics[width=0.5\textwidth]{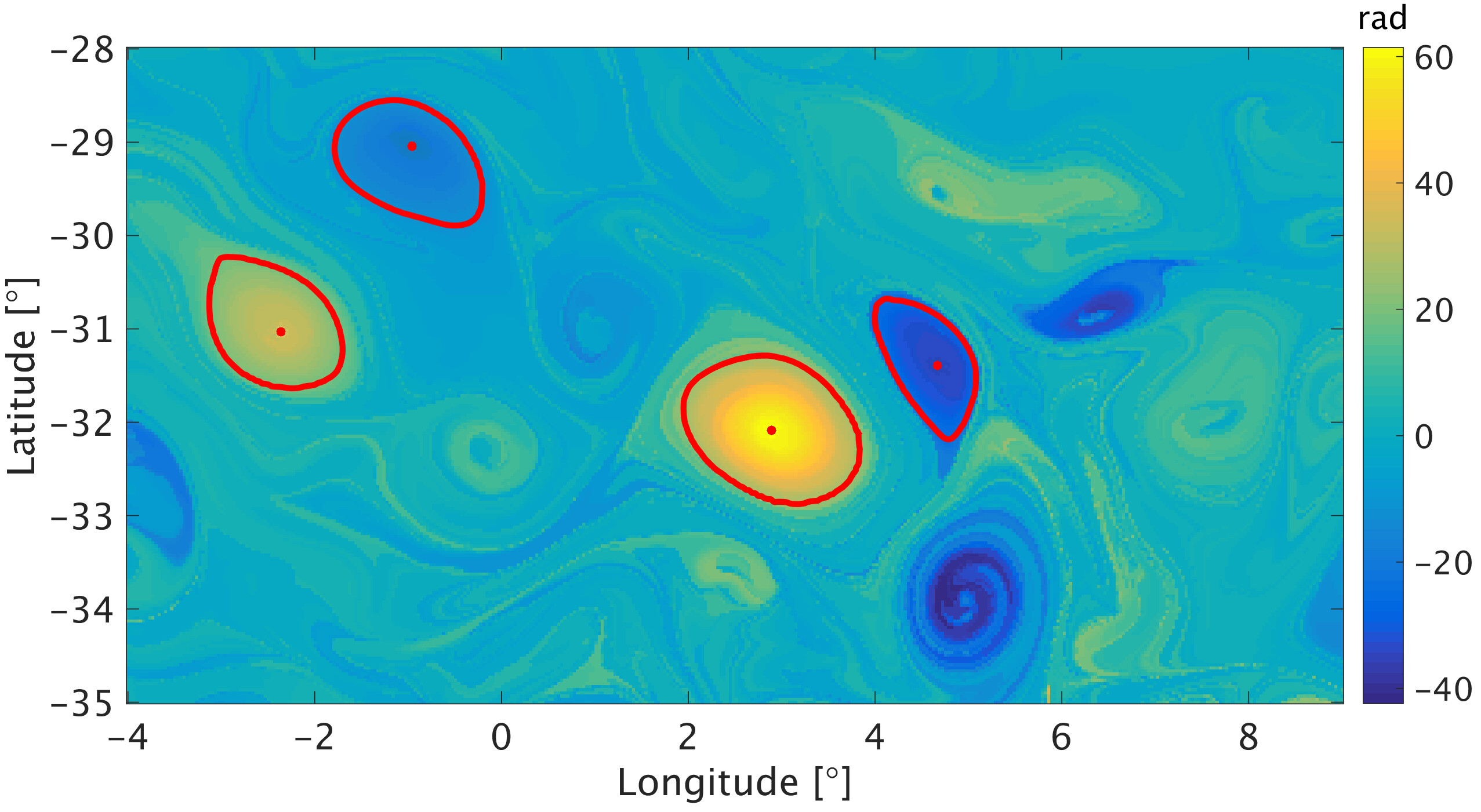}}
  \subfloat[]{\includegraphics[width=0.5\textwidth]{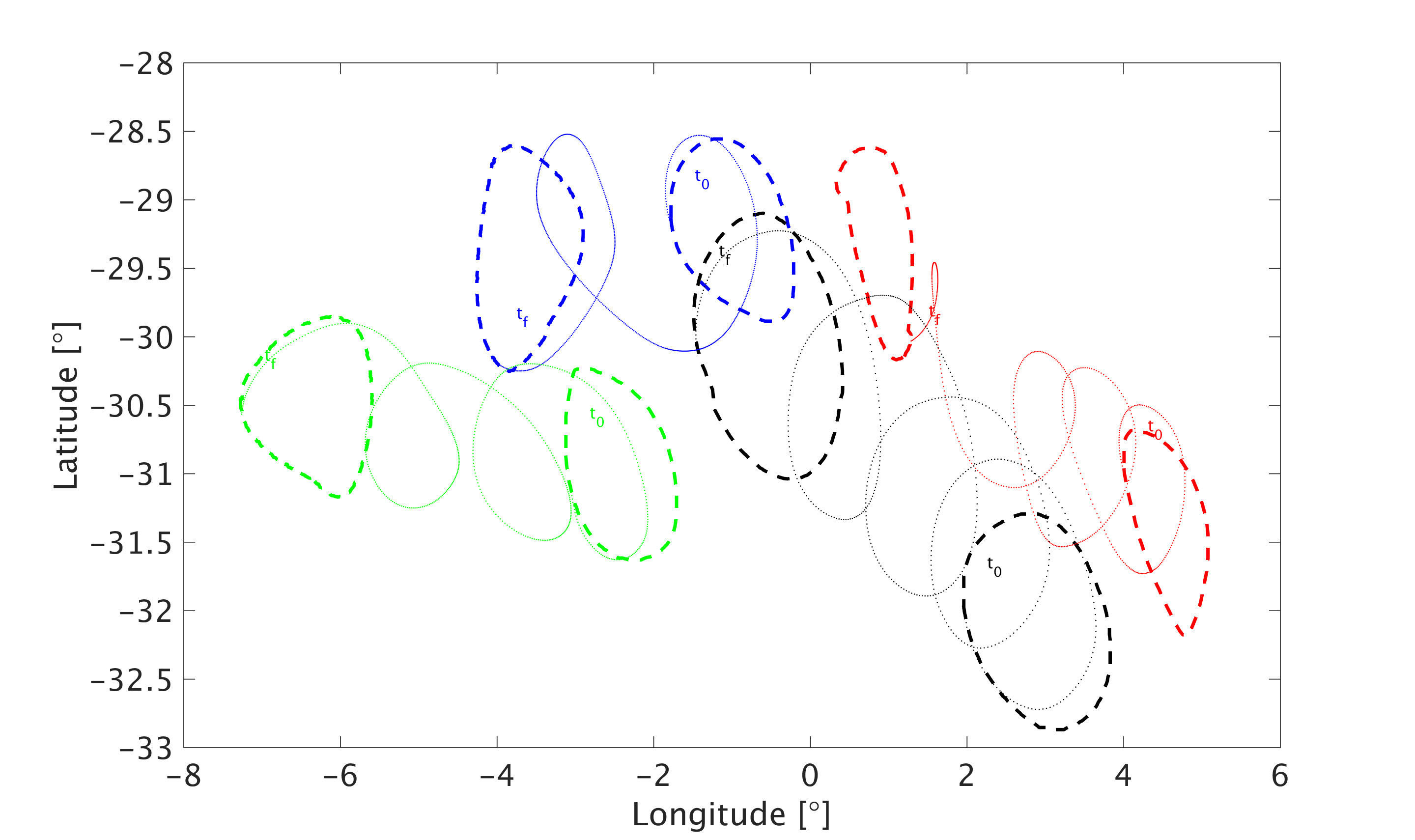}}

  \caption{a) Coherent eddies and their center at time $t_0$ extracted using definition.1 with the $\mathcal{LAM}_{t_f}^{t_0}$ map shown in background.  b) Their initial and final positions under Lagrangian advection along with trajectories of different particles initialized on their boundaries. (See the supplemental movie M2 for the complete advection sequence of these vortices. It also shows the oscillation of velocity vectors of different particles withing these eddies.)\label{fig:bis6}}
\end{figure}

The data used here fall withing the the Agulhas leakage region in the Southern Ocean, spans from $[-28^{\circ}N,-4^{\circ}W]$ and $[-35^{\circ}N,9^{\circ}W]$. It is well known for its long-lived eddies which carry and transport different water properties from the Indian ocean into the South Atlantic \cite{ruijter1999indian}. In this work, the time period between 11/11/2006 and 11/1/2007 is used. \footnotetext[2]{http://www.aviso.oceanobs.com/duacs/}

The AVISO data set (eq.\ref{eq:ssh}) is integrated over the period between $t_0=11$ November 2006 and $t_f =1$ January within an initial grid with step size $\Delta x_0=1/50^{\circ}$.
Fig.\ref{fig:bis6}(a) shows the $\mathcal{LAM}_{t_f}^{t_0}$ map in the background. Also it shows the boundaries and center of two couples of cyclone/anticyclone eddies extracted from the geostrophic velocity (eq.\ref{eq:ssh}).
Fig.\ref{fig:bis6}(b) shows their initial and final position under Lagrangian advection as well as trajectories of particles initialized on their boundaries. These eddies remain coherent all over the 3 months period of advection.  The complete advection sequence is illustrated in the movie M2.

\subsubsection{Three-dimensional Agulhas eddies in the Southern Ocean State Estimate model}

Here we also consider the Agulhas region, but we use a three-dimensional unsteady velocity field set obtained from the Southern Ocean state estimation (SOSE) model \cite{mazloff2010eddy}. 
Our experiment covers a period of $T = 30$ days, between $t_0 = 15$ May 2006 and  $t = 15$ June 2006. We select the domain defined by the longitudes $[11^{\circ}E, 16^{\circ}E]$, latitudes $[37^{\circ}S, 33^{\circ}S]$ and depths $[7, 2000]$ m.  We compute the three-dimensional $\mathcal{LAM}_{t_f}^{t_0}$  field by considering a uniform grid of $50\times50\times60$ points. 

\begin{figure*}[ht]
  \centering
  \subfloat[]{\includegraphics[width=0.4\textwidth]{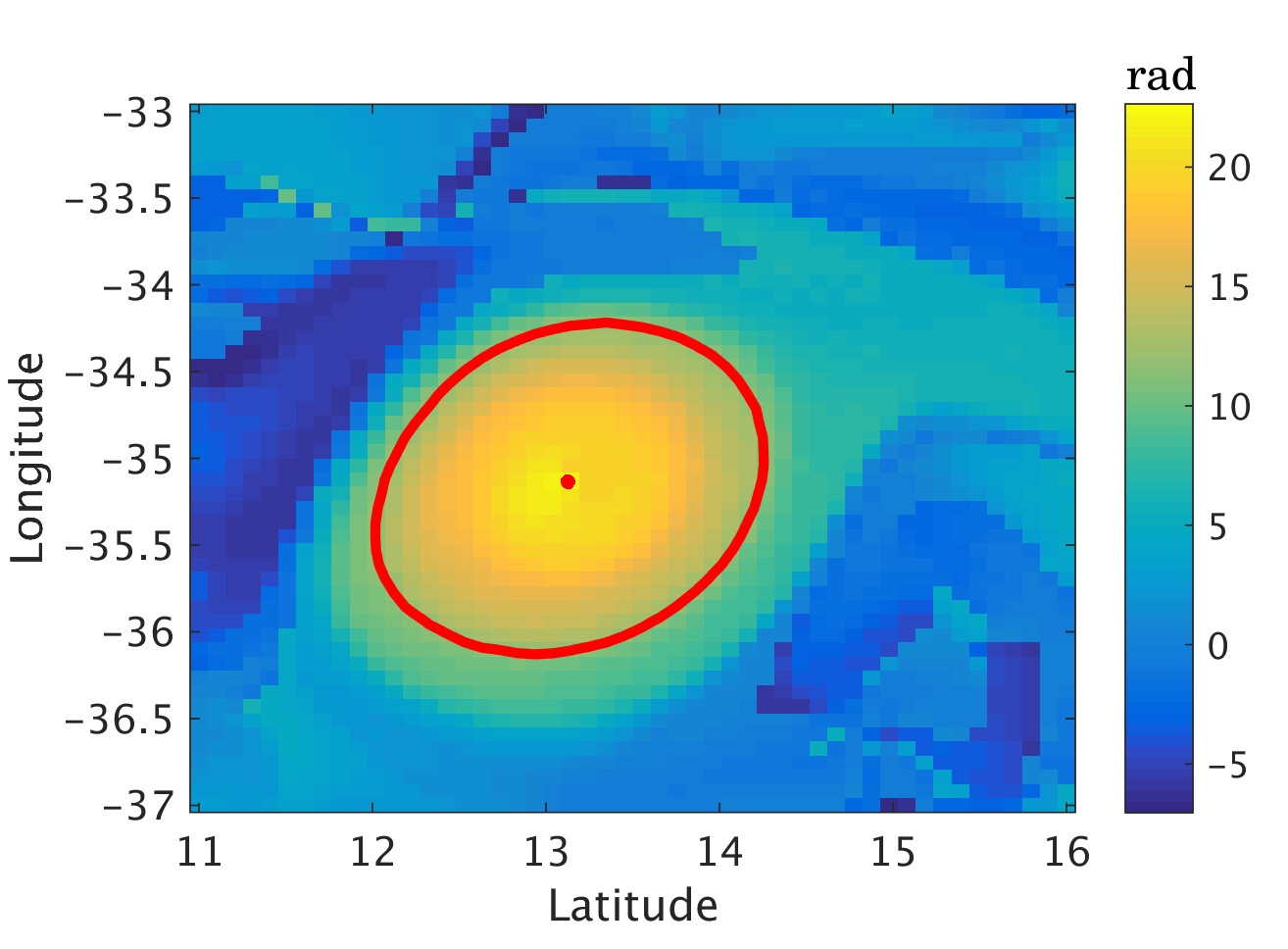}}
  \subfloat[]{\includegraphics[width=0.4\textwidth]{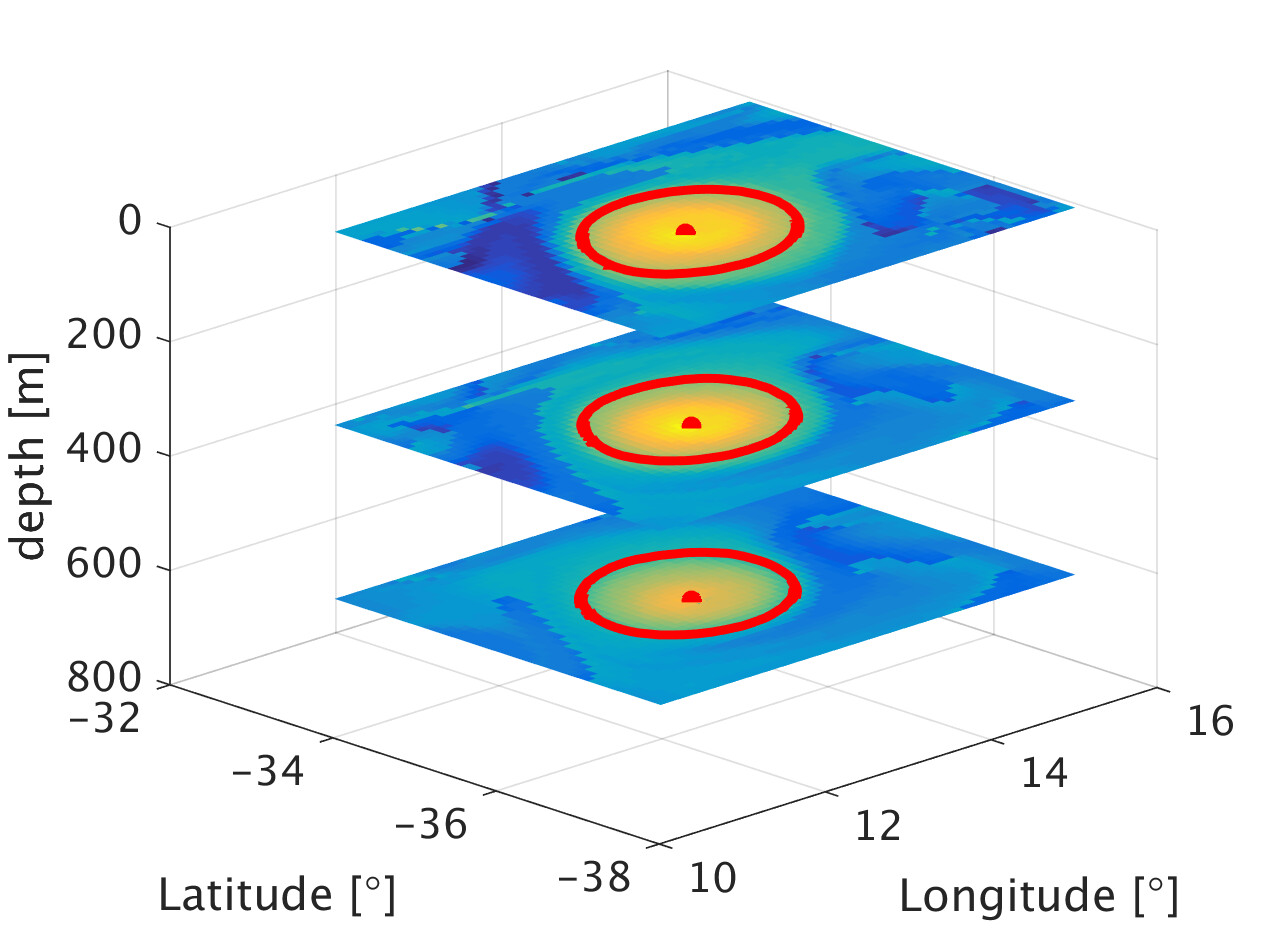}}

  \subfloat[]{\includegraphics[width=0.4\textwidth]{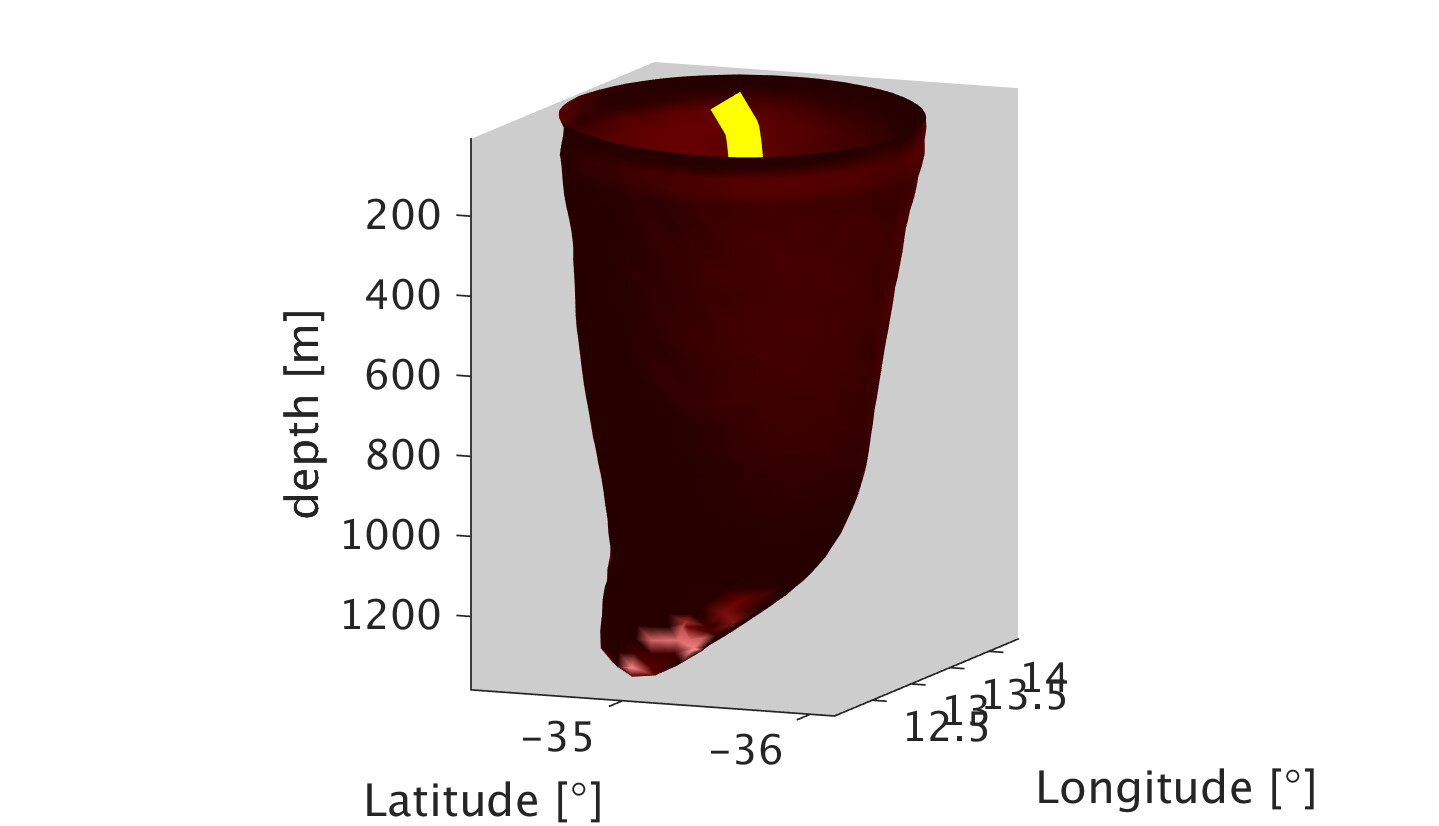}}
  \subfloat[]{\includegraphics[width=0.4\textwidth]{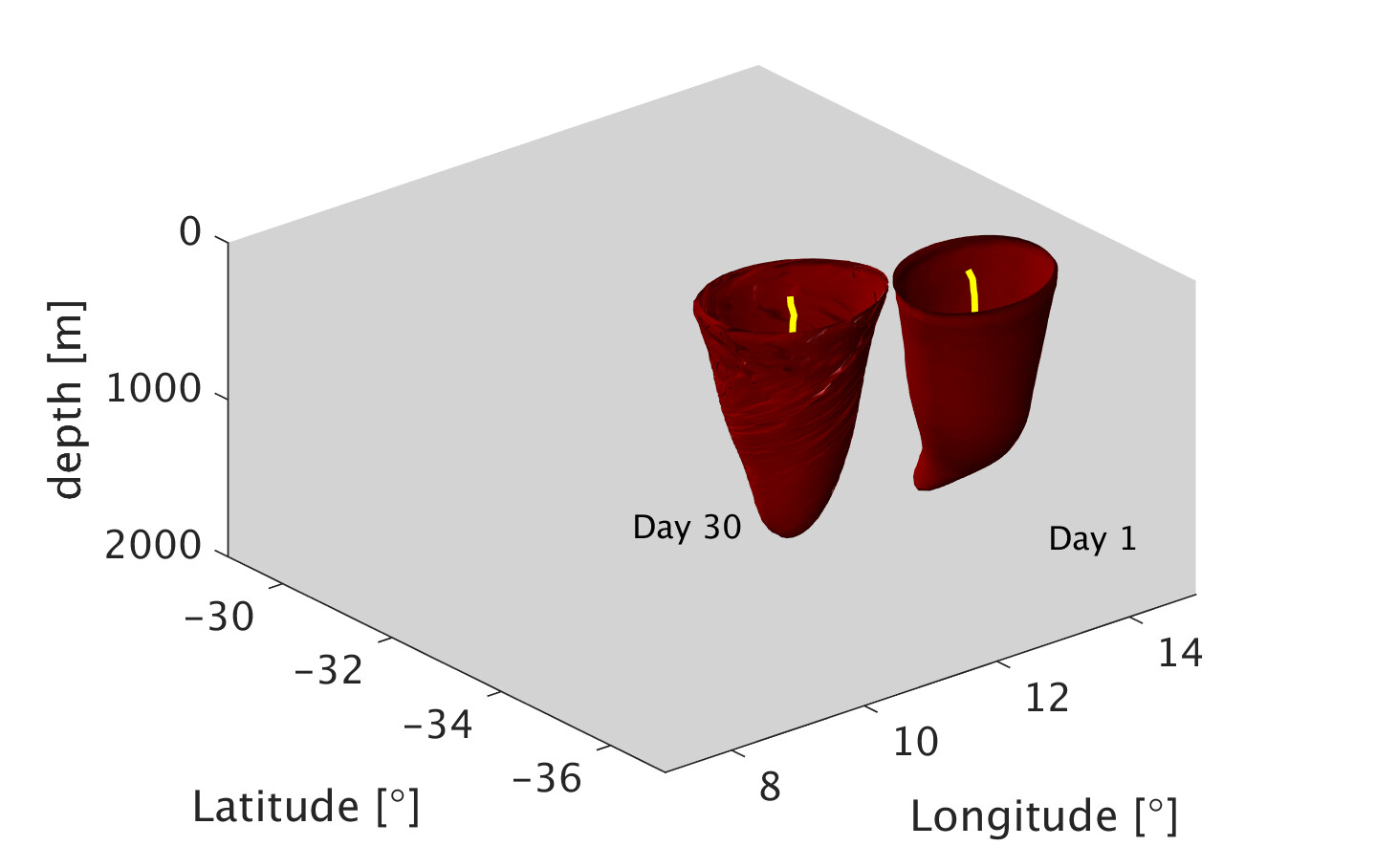}}

  \caption{a) Eddy's boundary and center extracted from the first layer of the three-dimensional field of the $\mathcal{LAM}_{t_f}^{t_0}$. b) Representative level surfaces of $\mathcal{LAM}_{t_f}^{t_0}$ showing the eddy's boundary and center at different depths. 
    c) Full view of the Lagrangian eddy boundary, the dark red surface is extracted using the definition.1, marking the vortex boundary for a mesoscale coherent Lagrangian eddy, extending from 10 m down to 1300 $m$ in depth. d) The initial and the  advected eddy's boundary 30 days later. (See the supplemental movie M3  for the complete advection sequence of this mesoscale eddy.)
\label{fig:bis7}}
\end{figure*}

Fig.\ref{fig:bis7}(a) shows in the background the first layer of the three-dimensional $\mathcal{LAM}_{t_f}^{t_0}$ field, the eddy's boundary and center.
Fig.\ref{fig:bis7}(b) shows a representative level surfaces of $\mathcal{LAM}_{t_f}^{t_0}$ field showing the eddy's boundary and center at different depths.
Fig.\ref{fig:bis7}(c) shows the full view of the coherent eddy (dark red) extending from 10 m down to 1300 m in depth. In the same image the eddy's center is presented in yellow. This eddy is extracted as level sets of $\mathcal{LAM}_{t_f}^{t_0}$.  Fig.\ref{fig:bis7}(d) shows the initial and the final position of the eddy after 30 days of Lagrangian advection. The complete three-dimensional advection sequence of this eddy is illustrated in the movie M3.

\section{Conclusion}

We proposed a new angular momentum based definition of coherent vortex as material domain in which particles exhibit similar intrinsic rotation.
This material domain is obtained from the total Lagrangian angle computed from the velocity vectors along their trajectories.
The performance of the proposed method is illustrated over different two- and three-dimensional flows. The findings show that proposed method is capable of extracting vortices boundaries and center in a sharp manner, and provides a complete dynamical evolution during their lifetime. Moreover, the present approach does not require advection of high grids, generally a very expensive computational procedure.

\bibliography{mybibfile}

\end{document}